\documentclass[aps,prl,superscriptaddress,12pt]{revtex4}
\usepackage{hyperref}

\usepackage{graphicx}

\usepackage{amsmath}
\usepackage{epsfig}

\begin{document}
\title{Experimental entanglement distillation of mesoscopic quantum states}

\author{Ruifang Dong}
\affiliation{Institut f\"{u}r Optik, Information und Photonik,
Max-Planck Forschungsgruppe, Universit\"{a}t Erlangen-N\"{u}rnberg,
G\"{u}nther-Scharowsky-Str. 1, 91058, Erlangen, Germany}

\author{Mikael Lassen}
\affiliation{Institut f\"{u}r Optik, Information und Photonik,
Max-Planck Forschungsgruppe, Universit\"{a}t Erlangen-N\"{u}rnberg,
G\"{u}nther-Scharowsky-Str. 1, 91058, Erlangen, Germany}
\affiliation{Department of
Physics, Technical University of Denmark, Building 309, 2800 Lyngby,
Denmark}
\author{Joel Heersink}
\affiliation{Institut f\"{u}r Optik, Information und Photonik,
Max-Planck Forschungsgruppe, Universit\"{a}t Erlangen-N\"{u}rnberg,
G\"{u}nther-Scharowsky-Str. 1, 91058, Erlangen, Germany}
\author{Christoph Marquardt}
\affiliation{Institut f\"{u}r Optik, Information und Photonik,
Max-Planck Forschungsgruppe, Universit\"{a}t Erlangen-N\"{u}rnberg,
G\"{u}nther-Scharowsky-Str. 1, 91058, Erlangen, Germany}
\author{Radim Filip}
\affiliation{Institut f\"{u}r Optik, Information und Photonik,
Max-Planck Forschungsgruppe, Universit\"{a}t Erlangen-N\"{u}rnberg,
G\"{u}nther-Scharowsky-Str. 1, 91058, Erlangen, Germany}
\affiliation{Department of Optics, Palack\'y University, 17.
listopadu 50, 77200 Olomouc, Czech Republic}
\author{Gerd Leuchs}
\affiliation{Institut f\"{u}r Optik, Information und Photonik,
Max-Planck Forschungsgruppe, Universit\"{a}t Erlangen-N\"{u}rnberg,
G\"{u}nther-Scharowsky-Str. 1, 91058, Erlangen, Germany}
\author{Ulrik L. Andersen}
\affiliation{Institut f\"{u}r
Optik, Information und Photonik, Max-Planck Forschungsgruppe,
Universit\"{a}t Erlangen-N\"{u}rnberg, G\"{u}nther-Scharowsky-Str.
1, 91058, Erlangen, Germany} \affiliation{Department of Physics,
Technical University of Denmark, Building 309, 2800 Lyngby, Denmark}

\date{\today}

\begin{abstract}
\textbf{The distribution of entangled states between distant parties
in an optical network is crucial for the successful implementation
of various quantum communication protocols such as quantum
cryptography, teleportation and dense
coding~\cite{Ekert91.prl,Bennett92.prl,Bennett92b.prl}. However,
owing to the unavoidable loss in any real optical channel, the
distribution of loss-intolerant entangled states is inevitably
inflicted by decoherence, which causes a degradation of the
transmitted entanglement. To combat the decoherence, entanglement
distillation, which is the process of extracting a small set of
highly entangled states from a large set of less entangled states,
can be used~\cite{bennett96.prl, kwiat01.nat,pan03.nat,eisert02.prl,
fiurasek02.prl, giedke02.pra, Opatrný00.pra, duan00.prl,
browne03.pra,fiurasek03.pra, ourjoumtsev07.prl}. Here we report on
the mesoscopic distillation of deterministically prepared entangled
light pulses that have undergone non-Gaussian noise. The entangled
light pulses~\cite{kumar,dong07.njp,dong08.ol} are sent through a
lossy channel, where the transmission is varying in time similarly
to light propagation in the atmosphere. By employing linear optical
components and global classical communication, the entanglement is
probabilistically increased. }
\end{abstract}

\maketitle

Entanglement distillation has been experimentally demonstrated for
spin $1/2$ (or qubit) systems exploiting a posteriori generated
polarization entangled states~\cite{pan03.nat}. However, the
implementation of a scheme that is capable of distilling
entanglement of continuous variable systems, where information is
encoded into mesoscopic carriers such as the quadratures of light
modes, has remained an experimental challenge. It has been shown
theoretically that if the wave function for the canonically
conjugate variables of the light mode is Gaussian, entanglement
distillation can be done only by utilizing highly non-linear (thus
difficult) operations no matter whether it is a pure or a mixed
state~\cite{eisert02.prl, fiurasek02.prl, giedke02.pra}. Several
protocols have been put forward~\cite{Opatrný00.pra, duan00.prl,
browne03.pra, fiurasek03.pra} and a proof of principle experiment on
the concentration of entanglement using {\it nonlocal} and
non-Gaussian operations has recently been
implemented~\cite{ourjoumtsev07.prl}.

In many practical scenarios, however, the transmitted quantum state
will be non-Gaussian: one example is the transmission of light
through a turbulent atmospheric channel, where the attenuation
coefficient will fluctuate in time, thus resulting in a non-Gaussian
quantum state~\cite{ursin07.natphy, book}. Fortunately, as we will
show in this letter it is possible to distill entanglement that has
undergone such noise by means of linear optical components, a simple
measurement induced Gaussian operation and classical communication.

We consider an optical field mode succinctly described by its
canonically conjugated quadrature amplitudes, which correspond to
the real and imaginary parts of the complex field. We denote
$\hat{X}$ the amplitude quadrature and $\hat{P}$ the phase
quadrature,The optical field can be represented by a
quasi-probability distribution known as the Wigner function $W(X,P)$
where $X$ and $P$ are eigenvalues of $\hat{X}$ and $\hat{P}$. Having
two optical fields, each described by the quadratures $(\hat{X}_A,
\hat{P}_A)$ and $(\hat{X}_B, \hat{P}_B)$, the joint state can be
described by the joint Wigner function $W(X_A,P_A,X_B,P_B)$. If this
function is Gaussian the joint optical state can be fully
characterized by its covariance matrix. For this case the
logarithmic negativity (which is an entanglement monotone), denoted
by LN, of the state is simply given by
\begin{equation}
LN=-Log_2\mu_{min},
\label{log}
\end{equation}
where $\mu_{min}$ is the smallest sympletic eigenvalue of the partial
transposed covariance matrix~\cite{vidal02.pra}.

\begin{figure}[h]
\includegraphics[width=0.9\textwidth]{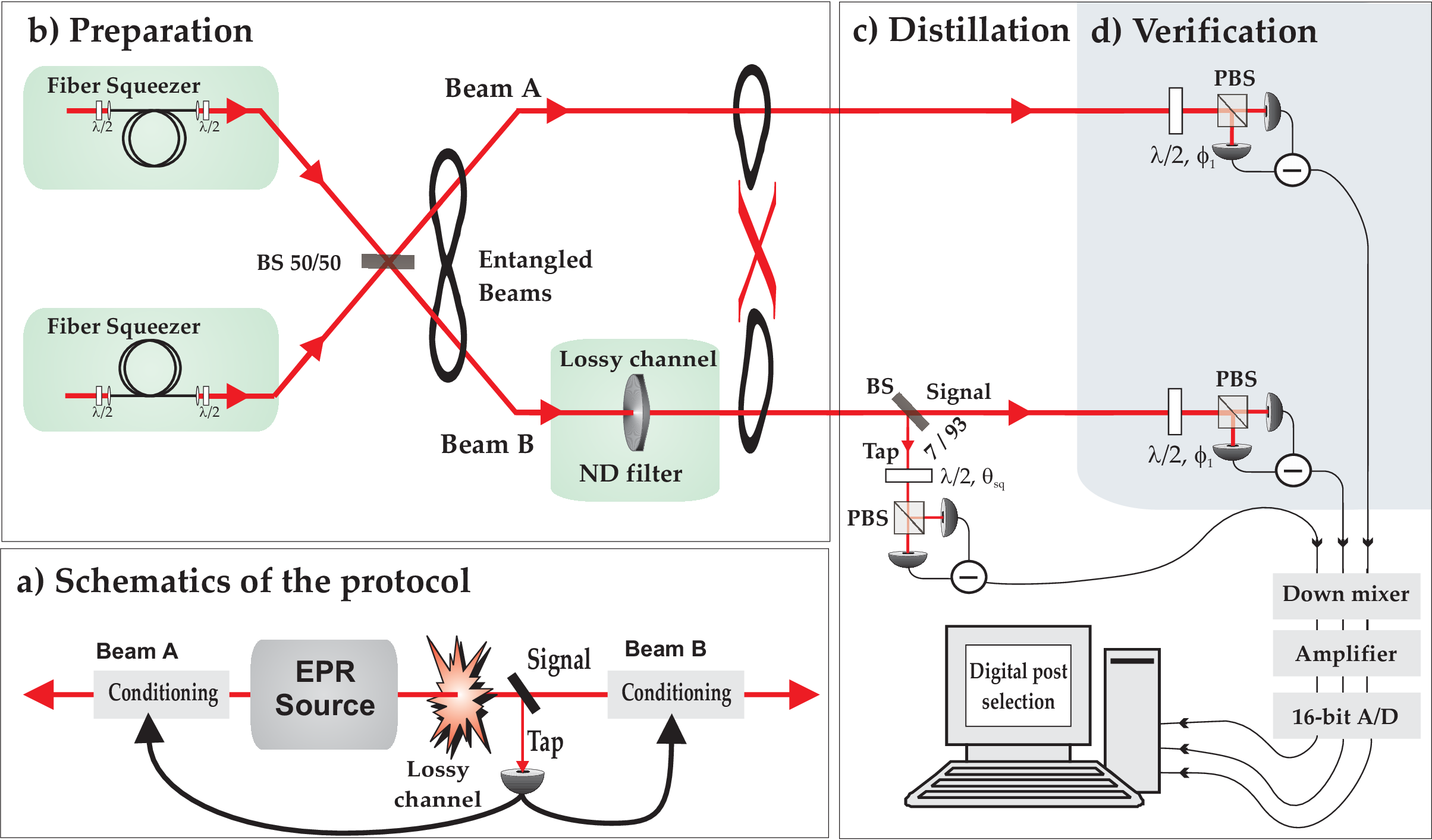}
\caption{\textbf{Schematics of the entanglement distillation
protocol and the experimental setup. a) A weak measurement on beam B
is diagnosing the state and subsequently used to herald the highly
entangled components of the state. b) The entangled states are
prepared by interfering two polarization squeezed beams on a 50/50
BS. The lossy channel is implemented by sending beam B through a
controllable neutral density filter (ND). c) The distillation
operation consists of a weak measurement (implemented by a 7\%
reflecting beam splitter) and homodyne measurement. d) The
verification measurement employs two independent polarization
detectors set to measured the conjugate quadratures $\hat{X}$ and
$\hat{P}$. The photocurrents are down-mixed, and digitized with a
fast A/D converter and fed into a computer. (see method section for
more details.)}} \label{schematics}
\end{figure}

Suppose now that one mode of the Gaussian entangled state is sent
through a medium with varying attenuation. We consider $N$ different
levels of attenuation. After the transmission the state turns into a
less entangled or even unentangled state and is described by the
convex mixture
\begin{equation}
W(X_A,P_A,X_B,P_B)=\sum_{i=1}^N p_i W'_i(X_A,P_A,X_B,P_B), \label{sum}
\end{equation}
where $p_i$ is the probability for a certain transmittance and the
Wigner function $W'_i$ represents the state in channel $i$ after
transmission. The individual constituents of the mixture are all
Gaussian functions but the sum is a non-Gaussian function. Because
of this non-Gaussianity, distillation can be enabled solely by
linear optics and feedforward as illustrated in
Fig.~\ref{schematics}. The operation consists of a weak measurement
(implemented by a 7\% reflecting beam splitter and a homodyne
detector measuring $\hat{X}$) followed by a probabilistic heralding
process where the remaining state is kept or discarded conditioned
on the measurement outcomes. If the outcome of the weak measurement
is larger than a specified threshold value, $X_{th}$, the remaining
state is kept~\cite{laurat03.prl, heersink06.prl, franzen06.prl,
suzuki06.pra, lance06.pra}, thus resulting in probabilistic recovery
of the entanglement with a corresponding increase in LN (see method
section). Note that our protocol cannot distill more entanglement
than is contained in the most entangled component of the mixture in
eqn.~(\ref{sum}). A notable difference between our distillation
approach and the schemes proposed in
Refs.~\cite{browne03.pra,fiurasek07.pra} is that our procedure
relies on single copies of distributed entangled states whereas the
protocols in Ref.~\cite{browne03.pra,fiurasek07.pra} are based on at
least two copies.

\begin{figure}[t]
\includegraphics[width=0.9\textwidth]{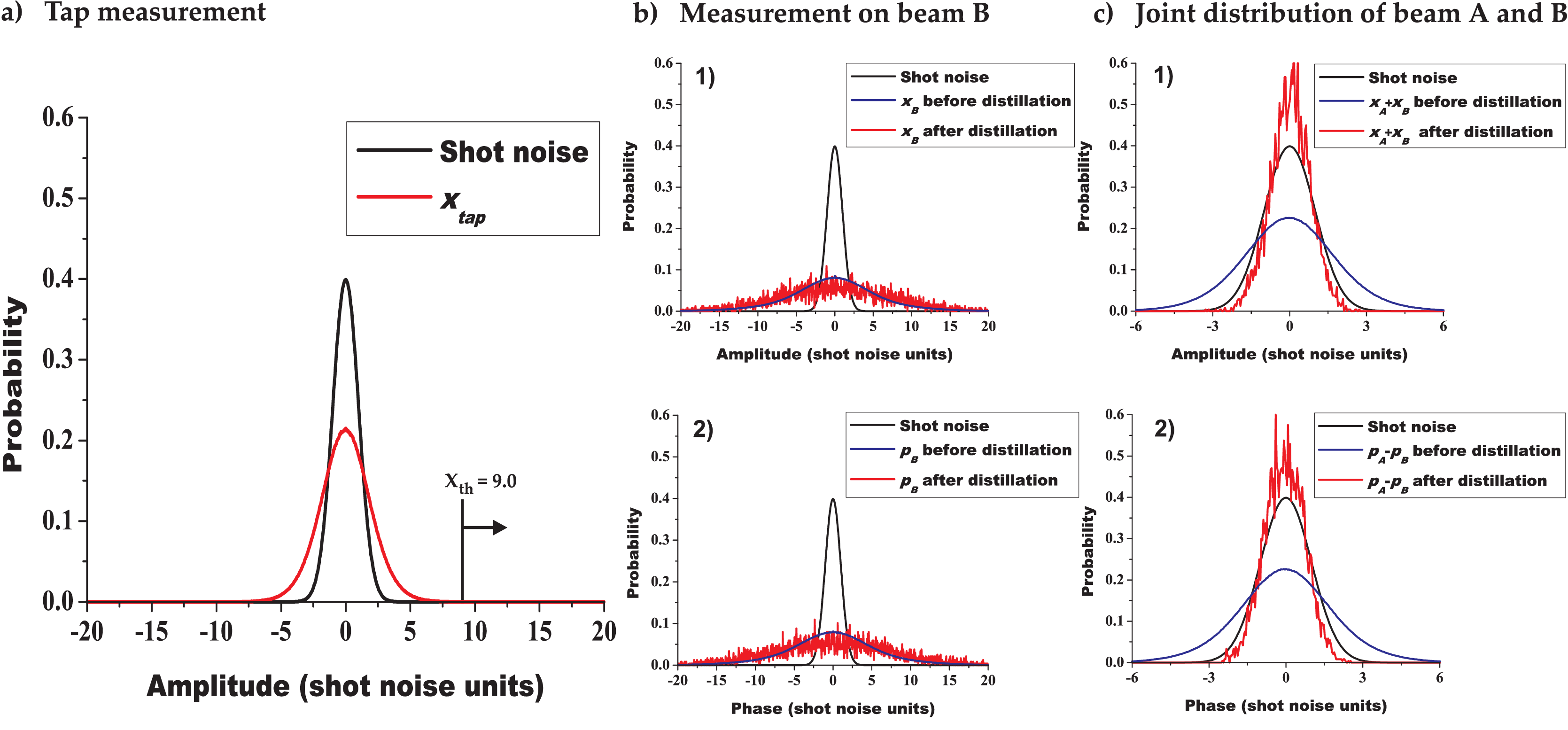}
\caption{(color online). \textbf{Experimentally measured marginal
distributions illustrating the effect of distillation. a) marginal
distribution for the amplitude quadrature in the tap measurement.
The vertical line indicates the threshold value chosen for this
realization. b) Marginal distributions associated with the
measurements of $X$ and $P$ of beam B (two left figures) and the
joint measurements $X_A+X_B$ and $P_A-P_B$ (two right figures). The
black, blue and red curves are the distribution for shot noise, the
mixed state before distillation and after distillation,
respectively.}} \label{entdistribution}
\end{figure}

The experimental realization is divided into three steps:
preparation, distillation, and verification as schematically
illustrated in Fig.~\ref{schematics}. The entangled states are
prepared by interfering two squeezed beams on a 50/50 beam splitter.
The squeezed beams are generated by exploiting the Kerr nonlinearity
experienced by ultra-short laser pulses in optical
fibers~\cite{dong07.njp}. To ease the detection process we produce
polarization squeezed states which inherently contain bright
polarization components that are used as local oscillators in
homodyne detection as described in ref.~\cite{kumar, dong08.ol}.
Details about the generation and measurement of entanglement can be
found in the method section. The Gaussian properties of the
entangled states are characterized by measuring the entries of the
covariance matrix, though assuming that the intra-correlations (such
as $\langle \hat{X}_A \hat{P}_A \rangle$) are zero due to the
symmetry of the states. From the covariance matrix we compute the
smallest sympletic eigenvalue from which we find the LN to be
$0.76\pm0.08$.

\begin{figure}[ht]
\includegraphics[width=0.7\textwidth]{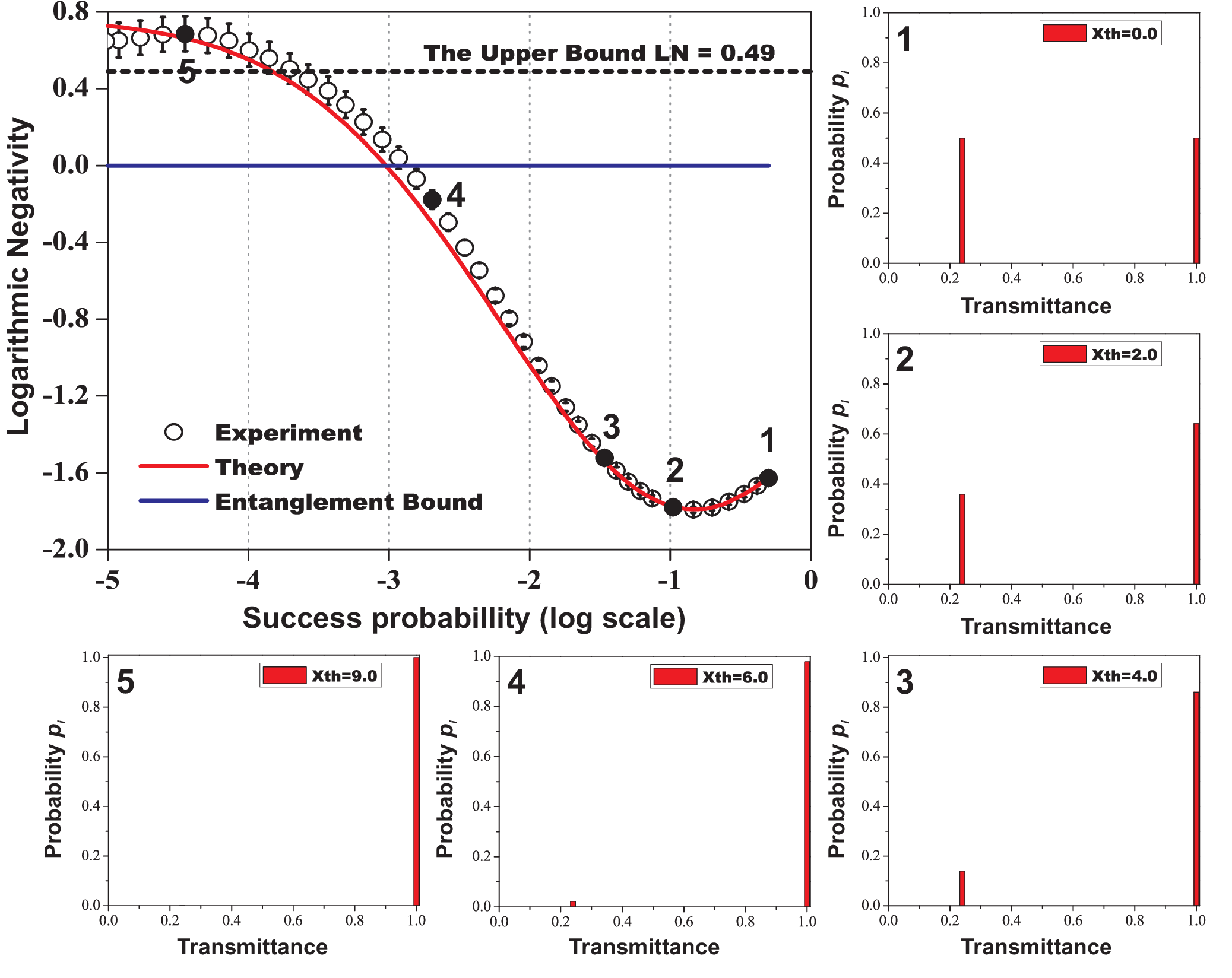}
\caption{(color online). \textbf{Experimental and theoretical
results outlining the distillation of an entangle state from a
discrete lossy channel. The experimental results are marked by
circles and the theoretical prediction is plotted by the red solid
line. The bound for Gaussian entanglement is given by the blue line,
and the upper bound for total entanglement before distillation is
given by the black dashed line. Both bounds are surpassed by the
experimental data. The weight of the two constituents in the mixed
state after distillation for various threshold values is also
experimentally investigated and shown in the plots labelled by 1-5.
The plots explicitly show the effect of the distillation. The error
bars depend partly on the measurement uncertainties mainly
associated with the finite resolution of the A/D converter, and
partly on the statistical uncertainties due to the finite
measurement time and the postselection process.}}
\label{Discrete_LN}
\end{figure}

\begin{figure}[h]
\includegraphics[width=0.7\textwidth]{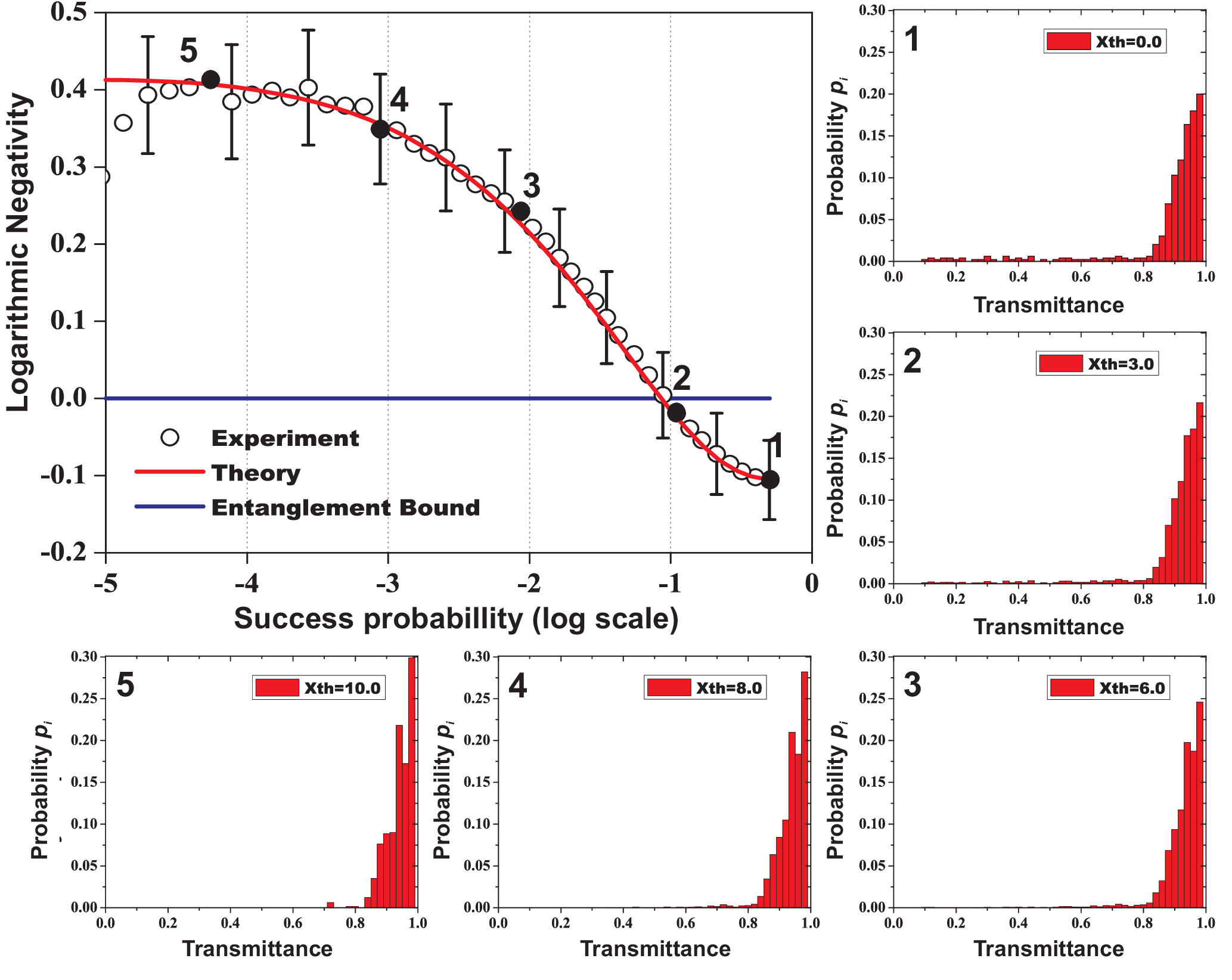}
\caption{(color online). \textbf{Experimental and theoretical
results outlining the distillation of an entangled state from a
semi-continuous lossy channel. The experimental results are marked
by circles and the theoretical prediction by the red solid curve.
The evolution of the mixture is directly visualized in the series of
probability distributions in 1-5. We see that for $X_{th}=10$ SNU
the probabilities associated with low transmission levels are 0 and
the probability for full transmission has increased to 30\% as
opposed to 20\% before distillation. It is thus clear that the
highly entangled states in the mixture have larger weight after
distillation. The theoretical prediction is given by the red curve.
}} \label{continuous}
\end{figure}

We implement the lossy channel by inserting a neutral density filter
with a variable transmittance in one of the entangled beams. The
entangled beam is then transmitted through a channel with $N=44$
different levels with corresponding transmittance from 0.1 to 1 in
steps of 0.9/45. Combining all these realizations of the experiment
a mixed state such as the one given by eqn.~(\ref{sum}) is formed
with the probabilities $p_i$ all being identical. However, after the
measurement we apply an envelope function to the probabilities which
allow us to change the probability amplitudes, thus implementing
different channels.
Distillation of entanglement is demonstrated for two different lossy
channels: First we consider a discrete channel where the
transmission randomly alternates between two different levels, and
secondly, we consider the semi-continuous channel where the
transmission alternates between 45 different levels with certain
probability amplitudes. The probability distributions of the
transmittance for the discrete channel and the continuous channel
are shown in Fig.~\ref{Discrete_LN}-1 and Fig.~\ref{continuous}-1,
respectively.


The discrete channel alternates between full transmission and 25\%
transmission each realisation occurring with a probability of 50\%.
After transmission in such a channel the resulting state is a
mixture of a highly and a weakly entangled state. For this state we
measure the Gaussian LN to be $-1.63\pm 0.02$.
The Gaussian entanglement is thus completely lost as a result of the
introduction of time-dependent loss.

The state is then fed into the distiller and we perform homodyne
measurements of beam A, beam B and the tap beam simultaneously. The
statistics of the quadrature measurements on the tap beam and beam B
as well as the joint distribution of $\hat{X}_A+\hat{X}_B$ and
$\hat{P}_A-\hat{P}_B$ are shown in Fig.~\ref{entdistribution}. From
the narrowing of the joint distributions to below that of the shot
noise, we conclude qualitatively that Gaussian entanglement has been
recovered.

The Gaussian LN has been computed for several choices of the
threshold value $X_{th}$, and is plotted in Fig.~\ref{Discrete_LN}
as a function of the associated success probabilities. Furthermore,
the probability coefficients of the two states in the mixture after
distillation are shown for different post-selection thresholds. Note
that as the threshold increases the mixture of the two Gaussian
states reduces to a single highly entangled Gaussian state, thus
demonstrating the act of Gaussification. Based on the experimental
parameters the theoretical predictions are computed and illustrated
in the figure by the red curve which is seen to be in good agreement
with the experimental results.

The results clearly show that the amount of Gaussian entanglement is
increased by the distillation operation. To estimate whether the
total entanglement is increased, we compute the upper bound for the
LN before distillation and verify that this bound can be surpassed
by the Gaussian LN after distillation (see the method section). The
upper bound of LN without the Gaussian approximation is computable
from the LN of each Gaussian state in the mixture~\cite{vidal02.pra}
and we find $LN_{upper}=0.49$, which is shown in
Fig.~\ref{Discrete_LN} by the dashed black line. We see that for a
success probability around $10^{-4}$ the Gaussian LN crosses the
upper bound for entanglement, and since the state at this point is
Gaussified we may conclude that the total entanglement of the state
has indeed increased as a result of the distillation.

Entanglement distillation also comes with a cost. As the degree of
entanglement is increasing, the number of distilled data or,
equivalently, the success probability, decreases. E.g. when the post
selection threshold is $X_{th}=9$ SNU, the Gaussian LN is
$0.67\pm0.09$ and the success probability is $1.69\times10^{-5}$.
The protocol has extracted only 8160 highly entangled states from a
total set of $2.4\times10^8$ less entangled states.


We now turn our attention to a communication channel which takes on
45 different transmission levels as opposed to the 2 level channel.
The distribution of the transmittance is illustrated in
Fig.~\ref{continuous}-1. This channel simulates a free-space optical
communications channel where atmospheric turbulence causes
scattering and beam pointing noise~\cite{book}. After propagation
through this channel the Gaussian LN of the mixed state is found to
be $-0.11\pm0.05$ which is substantially lower than the original
value of $0.76\pm0.08$. The state is subsequently distilled and the
change in the Gaussian LN as the threshold value increases (and the
success probability decreases) is shown in Fig.~\ref{continuous}. We
clearly see the trend that the entanglement available for Gaussian
operations is increased, ultimately reaching the level of
$LN=0.39\pm0.07$.



A summary of the measured values for the logarithmic negativities in
the various channels before and after distillation is presented in
table~\ref{table log-neg}. The demonstration of a distillation
protocol in this letter provides a crucial step towards the
construction of a quantum repeater~\cite{briegel98.prl} for
transmitting continuous variables quantum states over long distances
in channels inflicted by non-Gaussian noise. The various ingredients
for a continuous variable quantum repeater that could potentially
overcome non-Gaussian noise - a quantum
memory~\cite{julsgaard04.nat}, a teleportation
protocol~\cite{furusawa98.sci} and an entanglement distillation
protocol - have now all been experimentally realized and the next
step is to combine some of these technologies.

\begin{table}
\begin{tabular}{|l|c|c|c|c|}
  \hline
       ~~~~~~~~~~Channels   & ~~LN (before)~~  & ~~LN (after)~~ & Success rate $P_S$ ~~    \\ \hline
  ~~Perfect ~~           & $0.76\pm0.08$    & ~$0.76\pm0.08$       & ~1~                     \\
  ~~Discrete~~           & $-1.63\pm 0.02$  & ~$0.67\pm0.09$       & ~$1.69\times10^{-5}$~                     \\
  ~~Semi-continuous~~    & $-0.11\pm0.04$   & ~$0.39\pm0.07$       & ~$1.66\times10^{-5}$~                     \\  \hline
\end{tabular}
\caption{{\bf Measured Gaussian LN before and after distillation and
the corresponding success rates.}}\label{table log-neg}
\end{table}

\section*{METHODS}

\subsection*{Theory of distillation operation}
Theoretically, the distilled state reads
\begin{equation}
W_p(X_A,P_A,X_B,P_B)=\int_{X_{th}}^\infty\int_{-\infty}^\infty\sum_{i=1}^Np_i
W'_i(X_A,P_A,X_B,P_B)W_0(X_T,P_T)dX'_TdP'_T \label{post
measurement state}
\end{equation}
where $X'_B=\sqrt{T}X_B-\sqrt{1-T}X_T$,
$P'_B=\sqrt{T}P_B-\sqrt{1-T}P_T$, $X'_T=\sqrt{T}X_T+\sqrt{1-T}X_B$
and $P'_T=\sqrt{T}P_T+\sqrt{1-T}P_B$, and the Wigner function
$W_0(X_T,P_T)$ represents the vacuum mode entering the asymmetric
beam splitter with transmittance $T$. The question we now seek to
answer is whether this state is more entangled than the
pre-distilled state. The above mentioned measure - the Gaussian LN
(eqn.~(\ref{log})) - is only valid as an entanglement monotone if
the state is Gaussian, which is not the case for all stages of our
experiment. However, the Gaussian LN is a good measure of
entanglement useful for Gaussian operations. An example is
continuous variable quantum teleportation where an increase in the
Gaussian LN is directly linked with an increase in the teleportation
fidelity~\cite{adesso05.prl}. To prove that the total entanglement
of the state has increased as a result of distillation, we compute
an upper bound for the LN before distillation (given by
$LN=Log_{2}||\rho^T||_1$ where $\rho^T$ is the partial transposed
density matrix of the state) and demonstrate that this bound can be
surpassed by the Gaussian LN when the state is Gaussified after
distillation (as the measure then becomes exact).

\subsection*{Generation of entanglement}
The polarization squeezed beams are produced by launching two femto
second pulses with balanced powers onto the two orthogonal
polarization modes of two different fibers. We use two 13.2 m long
polarization-maintaining fibers. The pump source is a Cr$^{4+}$:YAG
laser at a wavelength of 1500~nm, repetition rate of 163 MHz. The
two orthogonal polarization modes in each fiber are squeezed and
temporally overlapped with a relative phase of $\pi/2$, thus
producing a circularly polarized light beam represented by the
Stokes parameter $S_3$. The relative phase is achieved and
controlled using an interferometric birefringence compensation and a
locking loop based on 0.1\% of the fiber output. Since the optical
excitation is along $S_3$, the orthogonal Stokes plane spanned by
the Stokes parameters $S_1$ and $S_2$ is "dark". However, it
contains a vacuum squeezed state which can be easily measured in a
polarization analyzer using the orthogonally polarized excitation.
Polarization squeezing in the "dark" plane is thus equivalent to
quadrature vacuum squeezing. We thus set
$\hat{S}(\theta_{sq})\rightarrow\hat{X}$ and $\hat{S}(\theta_{sq} +
\pi/2)\rightarrow\hat{P}$, where $\hat{S}(\theta_{sq})$ and
$\hat{S}(\theta_{sq} + \pi/2)$ are the "dark" Stokes operators of
the squeezing and anti-squeezing directions in the dark plane. The
two squeezed beams interfere on a 50/50 beam splitter to produce
entanglement and we subsequently measure all second order moments
between the quadratures using two homodyne detectors at
$17\pm0.5$~MHz.

\subsection*{Measurement of entanglement}
A general polarization, Stokes, measurement apparatus consists of a
half-wave plate, a polarizing beam splitter (PBS) and two intensity
detectors. The half-wave plate enables the measurement of different
Stokes parameters lying in the 'dark' plane when the light beam is
circularly ($S_3$) polarized. The PBS outputs are measured directly
by using intensity detectors with 98\% quantum efficiency InGaAs
photodiodes and with an incorporated low-pass filter in order to
avoid ac saturation due to the laser repetition oscillation. The
difference currents from the two detectors is produced and
subsequently mixed with an electronic local oscillator at 17 MHz,
low-pass filtered (1.9 MHz), amplified (FEMTO DHPVA-100) and finally
digitized by an A/D exit converter (Gage CompuScope 1610) at $10^7$
samples per second with a 16-bit resolution. After these data
processing steps, the noise statistics of the Stokes parameters are
characterized at 17~MHz relative to the optical field carrier
frequency with a bandwidth of 1~MHz. The signal is sampled around
this sideband to avoid the classical noise present in the frequency
band around the carrier. The measurement of polarization
entanglement is accomplished by applying identical polarization
measurements on beam A and B with the half-wave plates set to the
same angle (either measuring $\hat{X}$ or $\hat{P}$ in the "dark"
polarization plane). For each angle, the detected photocurrent noise
of beam A and B were simultaneously sampled $2.4\times10^{8}$ times
and the self and cross correlations between the data set could
easily be calculated. The covariance matrix was subsequently
determined and the logarithmic negativity was calculated.

\subsection*{Acknowledgments}
This work was supported by the EU project COMPAS (no. 212008), the
Deutsche Forschungsgesellschaft and the Danish Agency for Science
Technology and Innovation (no. 274-07-0509). ML and RF acknowledge
support from the Alexander von Humboldt Foundation and RF
acknowledges Measurement and Information in Optics (MSM6198959213),
LC 06007 of Czech Ministry of Education and 202/07/J040 of GACR.

\subsection*{Competing Interests}
The authors declare that they have no competing financial interests.

\subsection*{Correspondence}
Correspondence and requests for materials should be addressed to U.
Andersen (email: ulrik.andersen@fysik.dtu.dk).

\end{document}